# The 2008 outburst of the cataclysmic variable V358 Lyrae

Jeremy Shears, David Boyd, Tut Campbell, Shawn Dvorak, Robert Koff, Tom Krajci, Ian Miller, Gary Poyner, George Roberts and Arne Henden

**Abstract**

We report photometry of V358 Lyr during its 2008 November outburst, the first confirmed outburst since 1965. At its brightest the star was V=15.9 and the outburst amplitude was at least 7.3 magnitudes and lasted at least 23 days. The first 4 days of the outburst corresponded to the plateau phase and the star then faded at 0.13 mag/d over the next 7 days. There was then a drop in brightness to a temporary minimum at mag 19.5, which lasted less than 4 days, after which the star recovered to its previous brightness. The final stages of the outburst were poorly covered. Time resolved photometry during the outburst revealed no obvious large-scale modulations such as superhumps. Although some small apparently periodic signals were detected, their significance is uncertain. Our observations, and those of previous researchers, support V358 Lyr being a dwarf nova and are consistent with it being a member of the WZ Sge family.

**Introduction: History of V358 Lyr**

V358 Lyr was first identified as a variable star in the field of R Lyr by Hoffmeister of the Sonneberg Observatory in 1965, which he catalogued as S 9649 [1]. He found it present on two photographic plates: on Aug 4 it was $m_{pg}$ = 16 and had faded to $m_{pg}$ = 17 on Aug 19. By contrast, he reported it to be invisible on archival Palomar plates having a detection limit of $m_{pg}$ = 18. Thus he gave the photographic range as $m_{pg}$ 16 to fainter than 18 and suggested that it was a long period variable.

In an attempt to understand the behaviour of V358 Lyr, Galkina and Shugarov examined Moscow plates covering the 30 year interval from JD 2434112 to 2445264 (1952 Apr 9 to 1982 Oct 21) and found no trace of the star [2]. In a further investigation, Richter examined 310 Sonneberg plates covering JD 2438551 to 2446476 (1964 Jun 4 to 1986 Feb 14) and reported the star visible on only three [3]. Two of these were the plates exposed in 1965 Aug previously reported by Hoffmeister. Richter found no trace of the star to magnitude 21 on Palomar Sky Survey prints and suggested that V358 Lyr is either a classical nova or a dwarf nova of the WZ Sge family, but considered the latter more likely, not least because it is located so far from the galactic plane. The third plate identified by Richter, from 1965 Jun 29 when he reported the star to have $m_{pg}$ = 13.62, was recently re-examined in detail by Antipin, Samus and Kroll who concluded that the detection was most likely the result of a plate defect [4]. They presented the observations in Table 1 as being representative of the 1965 Jun outburst. They also examined a further 89 new plates in the Moscow collection covering JD 2445525 to 2450366 (1983 Jul 9 to 1996 Oct 9), but found nothing brighter than $m_{pg}$ 17. They concluded that V358 Lyr is most probably a faint cataclysmic variable with extremely rare outbursts.

We consulted the AAVSO International Database and found 2037 observations of V358 Lyr between JD 2449402 and 2454789 (1994 Feb 18 and 2008 Nov 11) [5]. However, the majority of these are negative observations and have limits around v = 15, brighter than the 1965 outburst, and so not a strong limit on subsequent outbursts. Over the past year or two, as CCDs have been used more frequently, the limits have been typically around V = 17.5 and no outbursts have been seen, leaving Hoffmeister's 1965 Aug event as the only hitherto recorded outburst of V358 Lyr.

**Discovery of the 2008 outburst**

The discovery circumstances of the 2008 outburst show how serendipity and international cooperation can play an important role in such events. GP and JS have used the Bradford Robotic Telescope (BRT) for several years to monitor for outbursts of unusual CVs and have found it especially useful for detecting and following up outbursts during periods of poor weather at their usual observing locations. The BRT is a remotely controlled public access facility located on Mount Teide, Tenerife, Canary Islands, Spain. It is operated by the University of Bradford, UK, and comprises a 0.35m SCT and an FLI MaxCam ME2 CCD camera. During 2008 Nov, the BRT had been out of commission as a result of damage to its weather station. The telescope was put back into service on the evening of Nov 22 and one of the first images taken was of the field of V358 Lyr. JS and GP examined the unfiltered image and noted that the star was apparently in outburst and, since only a single image was available from the BRT and it was cloudy at their observatories, they put out a request for confirmation on various internet email lists used by observers of CVs [6]. Confirmation that this was the first outburst of V358 Lyr since 1965 was received within a few minutes from SD as dusk had descended at his observatory in Florida, USA [7]. A Central Bureau for Astronomical Telegrams electronic circular was issued officially announcing the discovery on 2008 Nov 25 [8].

Photometry was obtained from the discovery and confirmation images using commercially available aperture photometry software with the V calibration sequence from the AAVSO Variable Star Plotter (VSP). Since these images were unfiltered (C, clear), we shall refer to the derived magnitudes as "CV"; some V-band measurements were also obtained:

| | | | |
|---|---|---|---|
| JS/GP discovery image from BRT | Nov 22.917 | 16.11 CV | +/- 0.04 |
| SD confirmatory image | Nov 22.98 | 16.14 CV | +/- 0.04 |
| TK image | Nov 23.08 | 15.87 V | +/- 0.03 |

Further analysis of filtered images taken on Nov 23.08 showed B-V = 0.09 +/- 0.05, consistent with a dwarf nova in outburst.

An image of V358 Lyr in outburst is shown in Figure 1a.

**Astrometric position**

The outburst allowed us to determine the position of V358 Lyr using a pre-release copy of the UCAC 3 as the reference catalogue:

BRT       RA 18h 59m 32.99s  Dec +42° 24' 12.0" +/- 0.2" (J2000)
SD        RA 18h 59m 32.97s  Dec +42° 24' 11.7" +/- 0.2" (J2000)
TK        RA 18h 59m 32.96s  Dec +42° 24' 12.0" +/- 0.1" (J2000)

These measurements are in good agreement with the position measured by Antipin *et al.* on Hoffmeister's discovery plate GC1387 [4]:

Antipin *et al.*  RA 18h 59m 32.95s  Dec +42° 24' 12.2" +/- 0.5" (J2000)

**Outburst light curve**

Figure 2 shows the light curve of the outburst based on the authors' observations plus data from the American Association of Variable Star Observers (AAVSO) International Database [5]. We shall frequently refer to dates in the truncated form JD = JD – 2454000. The form of the light curve is typical of a dwarf nova outburst. The first 4 days from the discovery of the outburst (JD 793 to 797) represents the plateau phase during which the star was virtually constant at around mag 16.2 to 16.3CV. Between JD 797 and 804 there was a decline to around mag 17.2 at 0.13 mag/d, typical of the rate of decline for a dwarf nova. Unfortunately rather few data exist after JD 804 properly to constrain the light curve in the later stages of the outburst, but it is evident that there was a remarkable dip to a temporary minimum of magnitude 19.5 on JD 806 followed by a recovery to 17.3 on JD 808. This temporary minimum was identified by only a single observation and therefore it is possible that a fainter level was achieved. The final positive observation was on JD 816, 23 days after the outburst was detected, after which we presume it returned to quiescence.

A deep image of the field of V358 Lyr on JD 831, 38 days after detection of the outburst, failed to record the star even though the limiting magnitude was around 21 (Figure 1b). Similarly, we found the star was undetected in two images we obtained at the USNO-Flagstaff during 2004:

        2004 Sep 3   1.0 m telescope     fainter than V = 22.2
        2004 Oct 15  1.55 m telescope   fainter than V = 22.5

John Thorstensen also found the star undetected in an image taken with the 2.4m Hiltner telescope at the MDM observatory on Kitt Peak during 2005, with a limiting magnitude of V= 23.2 [9].

Thus the outburst range was 15.9 V to fainter than 23.2 V, an amplitude of at least 7.3 magnitudes.

**Time resolved photometry**

The authors conducted unfiltered time-series photometry of V358 Lyr using the instrumentation shown in Table 2 and according to the observation log in Table 3. In all cases raw images were flat-fielded and dark-subtracted, before being analysed using aperture photometry software with the V calibration sequence as described in the previous section. Small systematic differences were apparent between observers, which were at least in part due to the different CCD cameras used. To overcome these differences the data were normalised by removing the mean and linear trends, as is common practise when the aim is to investigate periodic signals.

The resulting de-trended light curves are shown in Figures 3a to d, each covering 2.4 days of the outburst. Each photometry run was necessarily rather short due to the rather unfavourable position of the star in the north-western sky at the time of year that the outburst occurred. These curves show that there is considerable scatter in the data of up to +/- 0.05 mag. This is at least partly due to the inherent errors associated with each photometric measurement, as indicated by the error bars, resulting from the relative faintness of the star. However, there is also a suggestion that some of the variation could be a real variation in V385 Lyr, considering that in several of the runs the photometric errors are smaller than the actual variation. This could be due to flickering which has been observed during the outbursts of many dwarf novae.

It is clear from a cursory examination of the light curves that no large-scale modulations are present, such as superhumps. Nevertheless we performed a Lomb-Scargle analysis on the data using the Peranso software [10] to investigate whether any small-scale periodic signals are present. The resulting power spectrum, shown in Figure 4, has its highest peak at 0.07862(22) d (113.2 min) and a smaller peak at 0.05556(32) d (80.4 min). The superhump period error estimates are derived using the Schwarzenberg-Czerny method [11]. Several other statistical tests in Peranso, including ANOVA, CLEANest and DCDFT, gave the same result. To investigate the significance of the two signals, we employed 200 iterations of the False Alarm Probability (FAP) algorithm in Peranso, a Fisher randomisation test based on a Monte Carlo permutation procedure [12], and found the following values:

| Signal | FAP |
| --- | --- |
| 0.07862(22) d | 0.005 |
| 0.05556(32) d | 0.015 |

Normally an FAP <0.01 is considered to be significant. Thus the shorter period signal is unlikely to be significant, whereas it appears that the 0.07862(22) d signal is at least on the borderline of significance. However, we note that there are many other peaks in the original power spectrum and none of the signals we have discussed is particularly strong, so we hesitate to draw any conclusions about these signals.

**Discussion: V358 Lyr as a WZ Sge system**

Our observations, including the profile of the outburst light curve, its amplitude and the B-V colour index, are consistent with V358 Lyr being a dwarf nova. Given speculation by earlier researchers that V358 Lyr may actually be a dwarf nova of the WZ Sge family, it would be instructive to review the evidence for such a classification.

Dwarf novae are a type of cataclysmic variable star in which a cool main sequence secondary star loses mass to a white dwarf primary. Material from the secondary falls through the inner Lagrangian point and, because it carries substantial angular momentum, does not settle on the primary immediately but forms an accretion disc. From time-to-time, as material builds up in the disc, thermal instability drives the disc into a hotter, brighter state causing an outburst in which the star brightens by several magnitudes. Dwarf novae of the SU UMa family occasionally exhibit superoutbursts which last several times longer than normal outbursts and may be up to a magnitude brighter. During a superoutburst the light curve of a SU UMa star is characterized by superhumps. These are modulations which are a few percent longer than the orbital period [13]. They are thought to arise from the interaction of the secondary star orbit with a slowly precessing eccentric accretion disc. The eccentricity of the disc arises because a 3:1 resonance occurs between the secondary star orbit and the motion of matter in the outer accretion disc.

WZ Sge stars represent a sub-class of the SU UMa family that contains more highly evolved systems. These systems have very short orbital periods, long intervals between outbursts, typically decades, and exceptionally large outburst amplitudes, usually exceeding 6 magnitudes [13]. We shall consider whether V385 Lyr exhibits each of these characteristics.

Certainly, the outburst interval of 43 years between the two known outbursts of V358 Lyr is exceptionally long and considerably longer than the 33 years of the prototype, WZ Sge, itself. However, in spite of the archival research conducted by several people described earlier in this paper, covering some 56 years, it is possible that other outbursts have been missed.

V358 Lyr's outburst amplitude of at least 7.3 magnitudes is similar to several WZ Sge stars which typically have amplitudes in the range of ~ 6 to 8 magnitudes, including WZ Sge, HV Vir, EG Cnc and AL Com [14 to 17]. Moreover the outburst light curve we present is similar to other WZ Sge systems and we note a particular similarity to AL Com which has exhibited a remarkable dip, more properly called a temporary minimum, which interrupts the plateau phase during several outbursts [18-20]. For example, the temporary minimum during the 1995 superoutburst of AL Com occurred 28 days after maximum and lasted 3 to 4 days following which it returned to roughly the original brightness before entering the decline phase [17]. In the case of V358 Lyr, the duration of the dip is not well constrained, but was less than 4 days. The cause of such temporary minima is not known.

As noted above, WZ Sge systems have short orbital periods and they become more common at the lower end of the orbital period distribution of SU UMa systems, towards the period minimum (~78 min or 0.054 d [13]). Unfortunately our observations did not allow $P_{orb}$ to be measured, although it is tempting to speculate that one of the signals in the power spectrum reported above could be related to $P_{orb}$, but we need to stress that none of the signals are convincing. The shorter, less significant, period (0.05556(32) d) is within the range of $P_{orb}$ values of WZ Sge systems given in reference 20, but we note that if the strongest signal (0.07862(22) d) were $P_{orb}$, then V358 Lyr would have the longest $P_{orb}$ of currently known WZ Sge systems, just above RZ Leo ($P_{orb}$ = 0.07651 d [22]). During the preparation of this paper we became aware of a report from Taichi Kato *et al.* where a signal was reported at 0.05563(3) d during the 2008 outburst and which they interpreted as $P_{sh}$ [23]. Whilst this period is consistent with the shorter of the two periods we identified, Kato *et al.* pointed out that the signal was very weak.

Thus the above observations are consistent with a WZ Sge classification for V358 Lyr. However, spectroscopic confirmation would be required which, at the time of writing this paper, the authors are not aware was conducted at any stage of the outburst.

One notable feature of the outburst of V358 Lyr is the apparent absence of superhumps that are characteristic of superoutbursts of SU UMa (and therefore also WZ Sge) systems. One possibility, of course, is that this was a normal outburst as opposed to a superoutburst. It was at once thought that WZ Sge stars only have superoutbursts, which would clearly present a major problem for our hypothesis that V358 Lyr is such a system. However it now appears that a few WZ Sge systems, including RZ Leo, probably do show normal outbursts from time to time [24, 25]. One further curiosity is that temporary minima or rebrightening events, such as in the V358 Lyr outburst, have only been seen in superoutbursts. Therefore could this have been a superoutburst without superhumps? On the face of it this is an apparent contradiction since superhumps are usually considered diagnostic of superoutbursts. For a number of years researchers have studied the factors which determine whether certain dwarf nova superhump. A principle factor is the mass ratio, q, of the interacting binary stars, where q= $M_2/M_1$, and $M_2$ is the mass of the secondary star and $M_1$ the mass of the primary. Superhumps have been observed in systems with q ~ 0.35 down to ~ 0.035 [26]. At lower values of q, the secondary becomes too small to perturb the accretion disc. However, Joe Patterson has pointed out that the lower limit is not well characterised since the superhump period would be only marginally longer than $P_{orb}$ and thus hard to distinguish [27]. Moreover, such systems tend to be faint in quiescence hence rather few $P_{orb}$ values have been measured. We simply have no data to indicate whether V385 Lyr falls into this category and it is possible that V358 Lyr is unique. Hopefully further data will be forthcoming in future outbursts, but of course the next outburst may be a long time in coming.

## Conclusions

An outburst of V358 Lyr was observed during 2008 some 43 years after the only other recorded outburst. At its brightest the star was V=15.9 and the outburst amplitude was at least 7.3 magnitudes. The first 4 days corresponded to the plateau phase and the star then faded at 0.13 mag/d over the next 7 days. There was then a drop in brightness to a temporary minimum at mag 19.5, lasting less than 4 days, before the star recovered to its previous brightness. The final stages of the outburst were poorly covered and the last positive detection was made 23 days after detection. At quiescence the star is fainter than mag 22.5.

Time resolved photometry during the outburst revealed no obvious large-scale modulations such as superhumps. Although some small apparently periodic signals were detected at 0.07862(22) d and 0.05556(32) d, their significance is uncertain.

Our observations, and those of previous researchers, support V358 Lyr being a dwarf nova. We note that some characteristics, including the infrequent outbursts, large outburst amplitude and the profile of the light curve, including a temporary minimum, are consistent with a WZ Sge classification. However, in the absence of spectroscopic confirmation this remains speculation.

V358 Lyr is observed as part of the British Astronomical Association Variable Star Section's (BAA-VSS) Recurrent Objects Programme [28]. This programme was set up as a joint project between the BAA-VSS and *The Astronomer* magazine specifically to observe poorly studied eruptive stars of various types where outbursts occur at intervals of greater than 1 year. We urge observers to continue to monitor it for future activity which may bring further important information to light regarding this enigmatic object.

## Acknowledgements

The authors are indebted to John R. Thorstensen (Dept of Physics & Astronomy, Dartmouth College, USA) for making available his photometry of the field of V358 Lyr obtained during 2005, which was important in determining the limits of the outburst amplitude. We gratefully acknowledge the use of observations from the AAVSO International Database contributed by observers worldwide and access to the Bradford Robotic Telescope operated by the Department of Cybernetics, University of Bradford, UK. Finally we thank the referees whose comments have helped to improve the paper.

**Addresses:**
JS: "Pemberton", School Lane, Bunbury, Tarporley, Cheshire, CW6 9NR, UK [bunburyobservatory@hotmail.com]
DB: 5 Silver Lane, West Challow, Wantage, Oxon, OX12 9TX, UK [drsboyd@dsl.pipex.com]

TC: 7021 Whispering Pine, Harrison, AR 72601, USA
[jmontecamp@yahoo.com]
SD: Rolling Hills Observatory, Clermont, FL, USA
[sdvorak@rollinghillsobs.org]
RK:    980 Antelope Drive West, Bennett, CO 80102, USA
[bob@AntelopeHillsObservatory.org]
TK: CBA New Mexico, PO Box 1351 Cloudcroft, New Mexico 88317, USA
[tom_krajci@tularosa.net]
IM: Furzehill House, Ilston, Swansea, SA2 7LE, UK
[furzehillobservatory@hotmail.com]
GP: 67 Ellerton Road, Kingstanding, Birmingham, B44 0QE, UK
[Garypoyner@blueyonder.co.uk]
GR: 2007 Cedarmont Dr., Franklin, TN 37067, USA,
[georgeroberts@comcast.net]
AH: AAVSO, 49 Bay State Rd, Cambridge, MA 02138, USA
[arne@aavso.org]

**References**

1. Hoffmeister C., Astronomische Nachrichten, 289, 205 (1967)
2. Galkina M.P. Shugarov, S.Yu., 1985, Perem. Zvezdy 22, 225
3. Richter G., *IBVS*, 2971 (1986)
4. Antipin S.V., Samus N.N. and Kroll P., *IBVS*, 5522 (2004)
5. AAVSO, http://www.aavso.org
6. Shears J., cvnet-outburst message 2768 (2008)
   http://tech.groups.yahoo.com/group/baavss-alert/message/2768
7. Dvorak S., cvnet-outburst message 2772 (2008)
   http://tech.groups.yahoo.com/group/baavss-alert/message/2772
8. Henden A., Shears J., Poyner G. and Dvorak S., CBET 1582 (2008)
9. Thorstensen, J.R. Private communication (2009). At our request, John Thorstensen kindly measured the limiting magnitude of an image of the field of V358 Lyr taken with the 2.4m Hiltner telescope at the MDM observatory on Kitt Peak, USA, on 2005 Sep 14
10. Vanmunster T., Peranso, http://www.peranso.com
11. Schwarzenberg-Czerny A., Mon. Not. Royal Astron. Soc., 253, 198 (1991)
12. Press, W.H. *et al.*, Numerical Recipes: The Art of Scientific Computing, 2nd ed, Cambridge University Press, New York (1992)
13. Hellier C., Cataclysmic Variable Stars: How and why they vary, Springer-Verlag, 2001
14. Patterson J. et al., PASP, 114, 721 (2002)
15. Kato T., Sekine Y. and Hirata R., PASJ, 53, 1191 (2001)
16. Patterson J. *et al.*, PASP, 110, 1290 (1998)
17. Patterson J. *et al.*, PASP, 108, 749 (1996)
18. Howell S.B. *et al.*, AJ, 111, 2367-2378 (1996)
19. Ishioka R. *et al.*, A&A, 381, L41-L44 (2002)
20. Uemura M., IBVS, 5815 (2008)
21. Ritter H. and Kolb U., *Astron. and Astrophys.*, **404**, 301 (2003) and available at http://www.mpa-garching.mpg.de/Rkcat/
22. Mennickent R.E. and Tappert C., A&A, 372, 563-565 (2001)
23. Kato T. *et al.*, PASJ, submitted (2009)

24. O'Donoghue D.O. *et al.*, MNRAS, 250, 363-372 (1991)
25. Osaki Y., PASJ, 47, 47-58 (1995)
26. Patterson J. *et al.*, PASP, 117, 1204-1222 (2005)
27. Patterson J. (2008). Communication posted on the "cba-chat" email list.
28. Poyner G., *J. Brit. Astron. Assoc.*, **106**, 155 (1996)

| Plate | Date in 1965 | JD | $m_{pg}$ |
|---|---|---|---|
| Te$_4$ 4601 | Jun 25 | 2438937.483 | >14.5 |
| GB 1905 | Jun 25 | 2438937.503 | >17 |
| GB 1908 | Jun 28 | 2438940.505 | >17 |
| GB 1911 | Jun 29 | 2438941.506 | >17 |
| GC 1387 | Aug 4 | 2438977.480 | 16.42 |
| GC 1388 | Aug 19 | 2438992.399 | 17.31 |
| GC 1389 | Aug 23 | 2438996.412 | >18.5 |

**Table 1: Sonneberg observations of V358 Lyr**
Data are from reference 3

| Observer | Telescope | CCD |
|---|---|---|
| DB | 0.35 m SCT | Starlight Xpress SXV-H9 |
| TC* | 0.3 m SCT | SBIG ST-9XE |
| SD | 0.25 m SCT | SBIG ST-9XE |
| RK | 0.25 m SCT | Apogee AP-47 |
| TK | 0.28 m SCT | SBIG ST-7E |
| IM | 0.35 m SCT | Starlight Xpress SXVF-H16 |

**Table 2: Instrumentation used**
* TC's raw photometry data were reduced by GR

| Date in 2008 (UT) | Start time (JD-2454000) | Duration (h) | Observer |
|---|---|---|---|
| Nov 23 | 793.527 | 2.8 | TK |
| Nov 23 | 794.493 | 2.5 | SD |
| Nov 24 | 794.537 | 2.4 | TK |
| Nov 24 | 795.242 | 3.3 | DB |
| Nov 24 | 795.245 | 3.5 | IM |
| Nov 25 | 795.504 | 1.4 | TC |
| Nov 25 | 795.528 | 2.4 | RK |
| Nov 26 | 796.294 | 2.0 | IM |
| Nov 27 | 797.509 | 2.5 | TC |
| Nov 28 | 798.342 | 0.8 | IM |
| Nov 30 | 800.530 | 1.6 | TK |
| Dec 1 | 801.535 | 2.4 | TK |
| Dec 2 | 802.536 | 1.9 | TK |
| Dec 3 | 804.532 | 2.2 | TK |

**Table 3: Log of time-series observations**

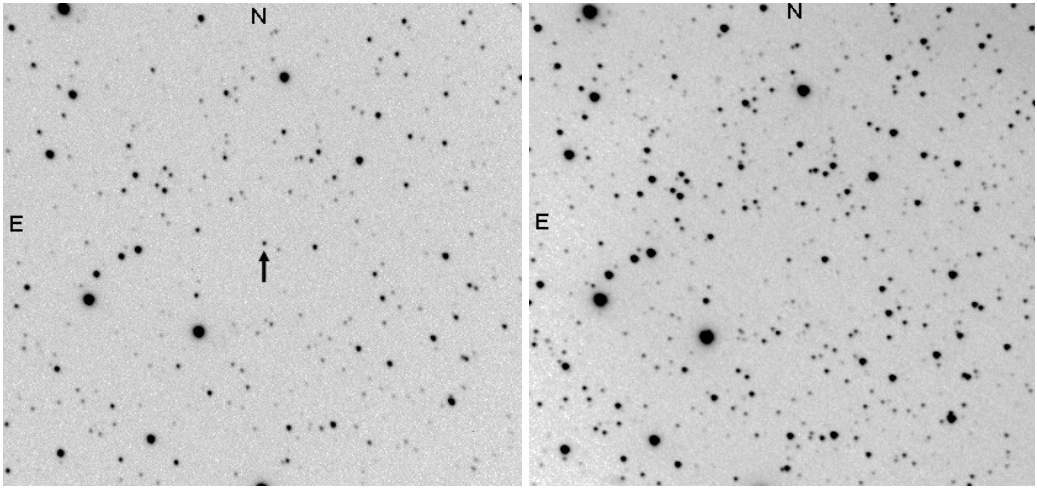

**Figure 1: Field of V358 Lyr**
(a) – left – 2008 Nov 24, in outburst at 16.2 CV; (b) – right – 2008 Dec 30, deep field with limiting magnitude fainter than 21 *(David Boyd)*

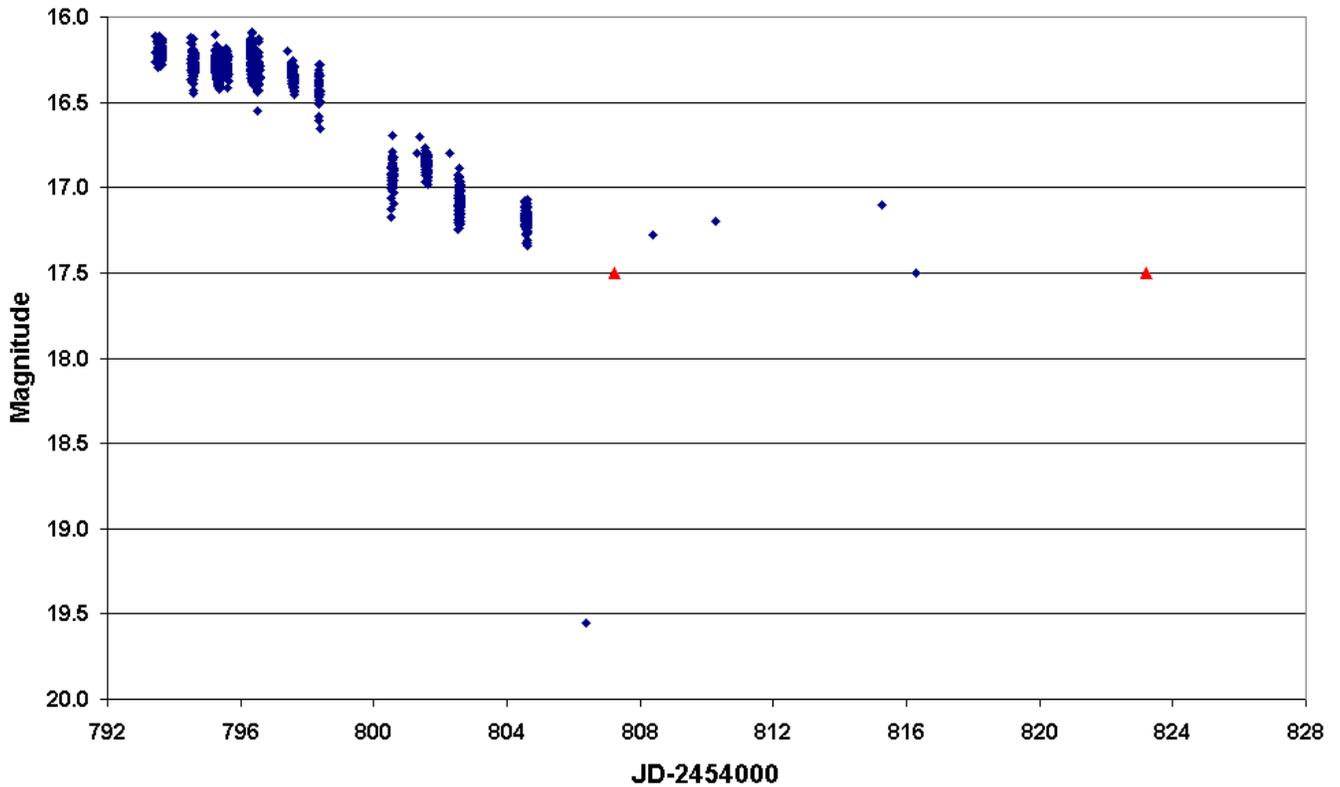

**Figure 2: Light curve of the outburst**
Note that the red triangles denote upper limit (i.e. "fainter than") observations

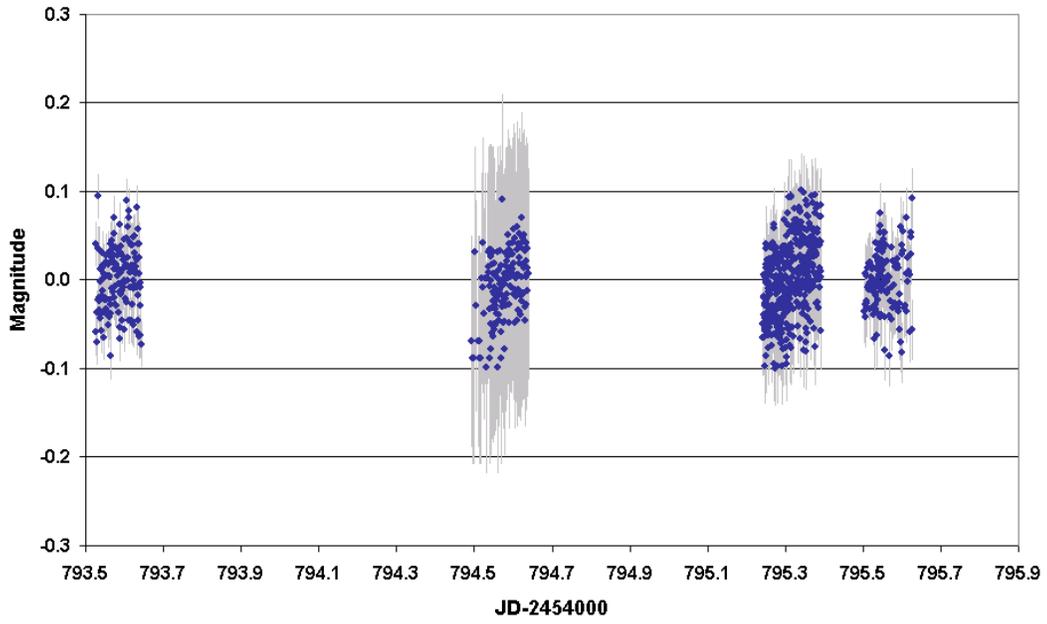
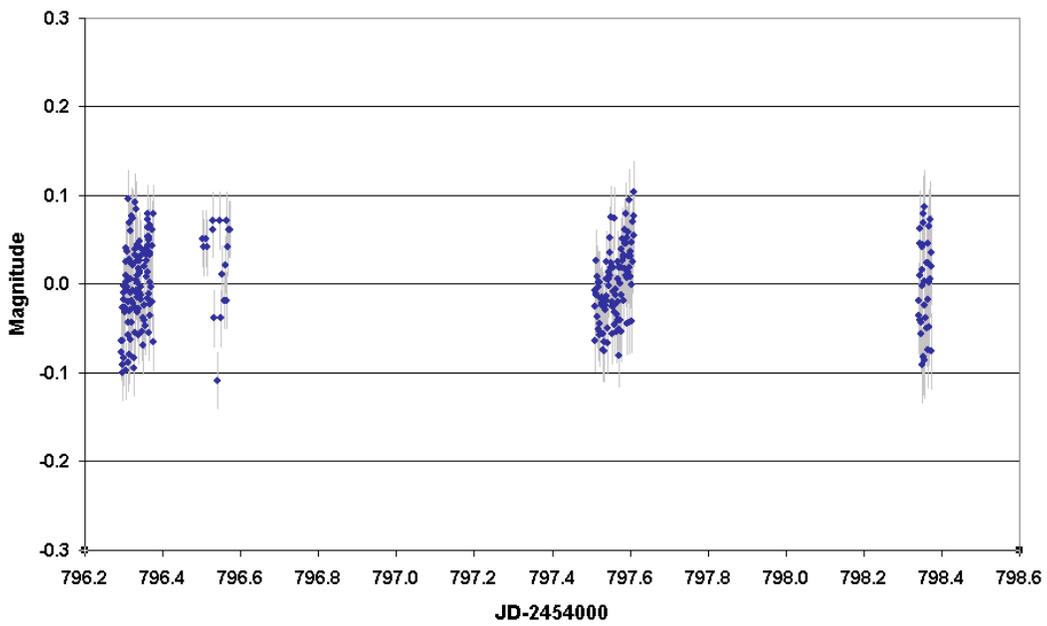

**Figure 3a and b: De-trended light curve of the outburst**

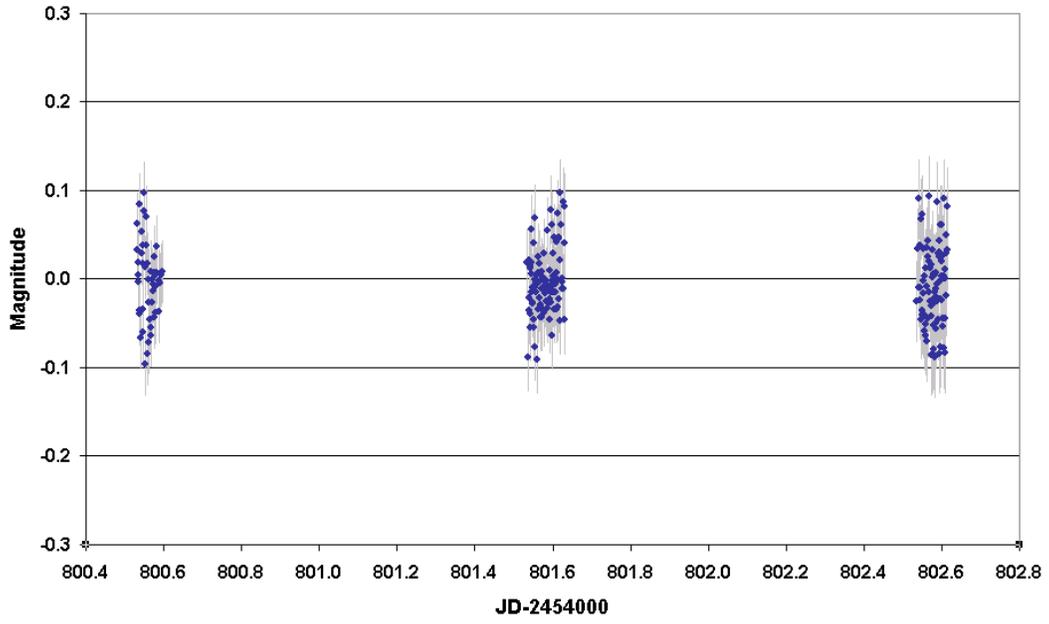
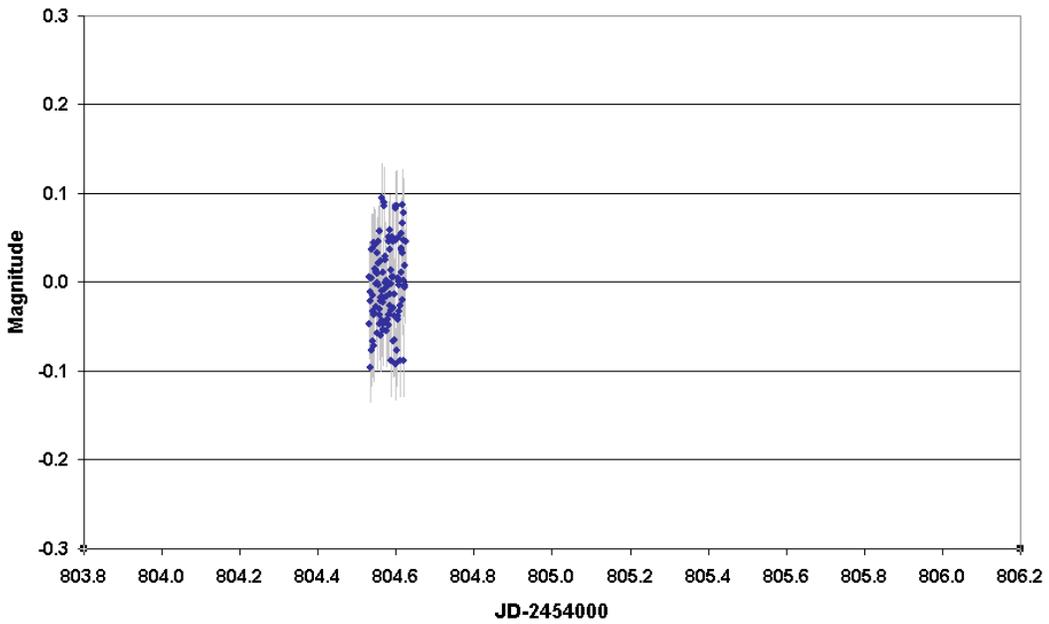

**Figure 3c and d: De-trended light curve of the outburst**

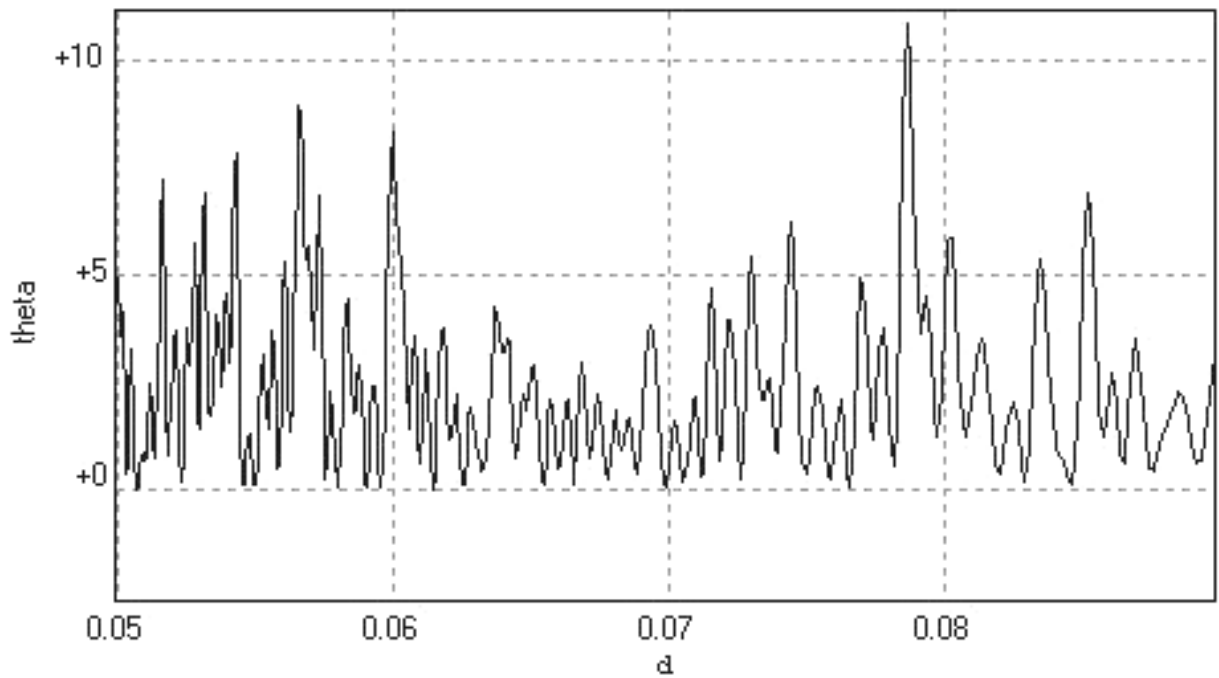

**Figure 4: Power spectrum of combined time-series data**